\pdfoutput=1
\RequirePackage{ifpdf}
\ifpdf 
\documentclass[pdftex]{sigma}
\else
\documentclass{sigma}
\fi

\newcommand{\brabra}{{\langle\!\langle}}
\newcommand{\ketket}{{\rangle\!\rangle}}
\newcommand{\C}{{\mathbb C}}

\newcommand{\percent}[1]{ {\%\%\%\%\%} }

\newcommand{\q}[1]{\left[ #1 \right]}
\newcommand{\Z}{{\mathbb Z}}
\newcommand{\nc}{\newcommand}
\nc{\rnc}{\renewcommand}
\nc{\nn}{\nonumber}
\newcommand{\bra}{{\langle}}
\newcommand{\ket}{{\rangle}}
\newcommand{\Log}{\operatorname{Log}}
\newcommand{\Aux}{\operatorname{Aux}}
\newcommand{\blt}{{\bullet}}

\begin{document}

\allowdisplaybreaks

\renewcommand{\PaperNumber}{081}

\FirstPageHeading

\ShortArticleName{Entanglement Properties of a Higher-Integer-Spin AKLT Model}

\ArticleName{Entanglement Properties of
a Higher-Integer-Spin\\ AKLT Model with Quantum Group Symmetry}

\Author{Chikashi ARITA~$^\dag$ and Kohei MOTEGI~$^\ddag$}

\AuthorNameForHeading{C.~Arita and K.~Motegi}

\Address{$^\dag$~Institut de Physique Th\'eorique CEA,
F-91191 Gif-sur-Yvette, France}
\EmailD{\href{mailto:chikashi.arita@cea.fr}{chikashi.arita@cea.fr}}

\Address{$^\ddag$~Okayama Institute for Quantum Physics,
Kyoyama 1-9-1, Okayama 700-0015, Japan}
\EmailD{\href{mailto:motegi@gokutan.c.u-tokyo.ac.jp}{motegi@gokutan.c.u-tokyo.ac.jp}}

\ArticleDates{Received July 06, 2012, in f\/inal form October 23, 2012; Published online October 27, 2012}

\Abstract{We study the entanglement properties of a higher-integer-spin
 Af\/f\/leck--Kennedy--Lieb--Tasaki model
 with quantum group symmetry in the periodic boundary condition.
We exactly
calculate the f\/inite size correction terms of the entanglement entropies
from the double scaling limit.
We also evaluate the geometric entanglement, which serves as another
measure for entanglement. We f\/ind
the geometric entanglement reaches its maximum at the isotropic point,
and decreases with the increase of the anisotropy.
This behavior is similar to that of the entanglement
entropies.}


\Classification{17B37; 81V70; 82B23}

\section{Introduction}

\begin{figure}[t]\centering
\includegraphics[width=70mm]{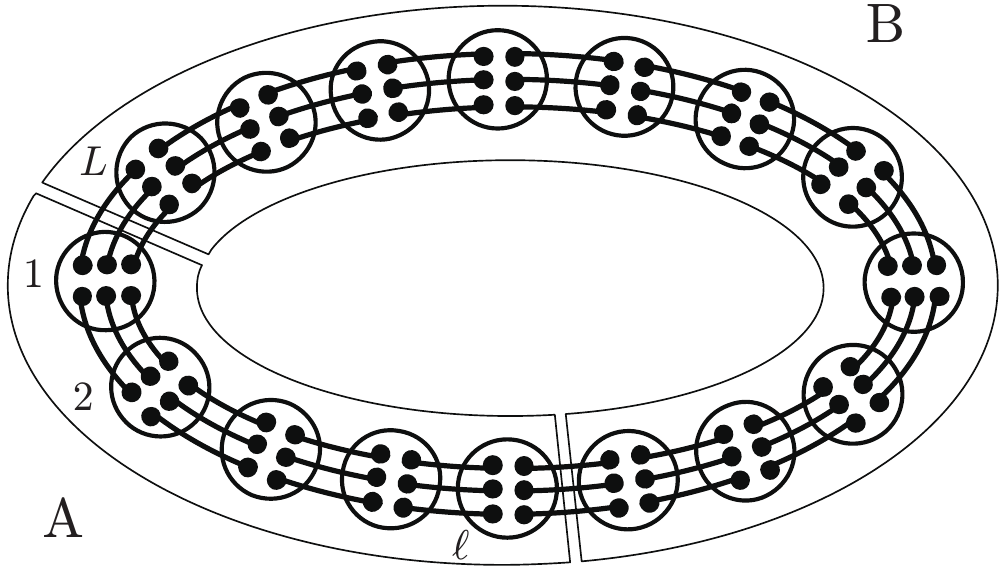}
\qquad
\includegraphics[width=70mm]{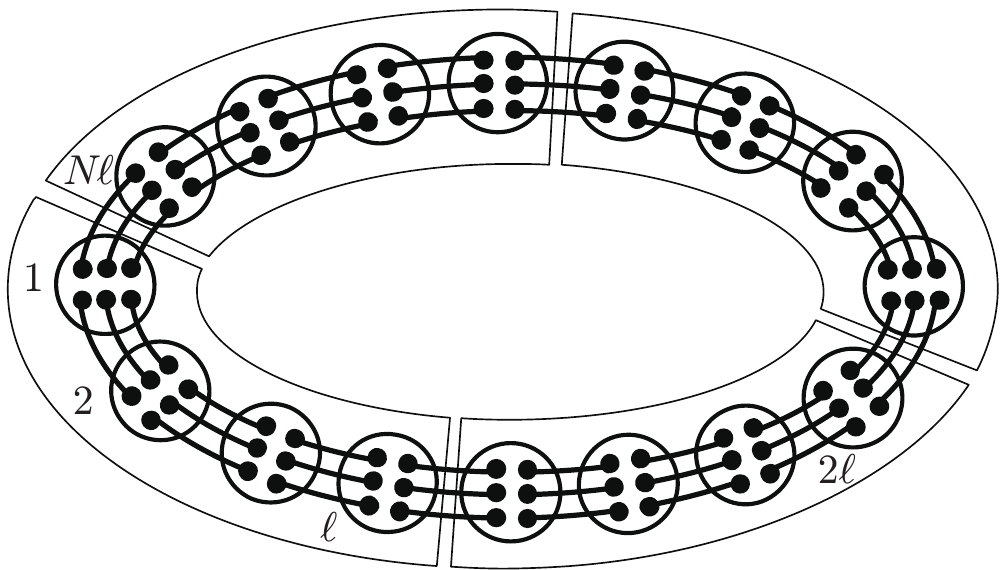}

\caption{Schematic pictures of
the entanglement entropies (left) and
the geometric entanglement (right)  for the $q$-AKLT model
with $S=3$.}
\label{fig:qVBS-EE-GE}
\end{figure}

Quantum entanglement is a fundamental feature in quantum mechanics,
and is a primary resource in quantum communication and quantum
computation \cite{BD,GMC,Ll,VMC}.
Entanglement has become an important tool to
characterize quantum many-body systems
(see \cite{AFOV} for example for a review).
In one dimensional spin systems,
typical quantif\/ications of quantum entanglement are
the R\'{e}nyi entropy $S_{\rm R} (L,\ell)$
and von Neumann entropy $ S_{\rm vN} (L,\ell)$
 of  a subsystem A with $\ell$ sites and environment
 B with $L-\ell$ sites (see  Fig.~\ref{fig:qVBS-EE-GE})
\begin{gather*}
   S_{\rm R} (L,\ell)=
 \frac{\log \operatorname{Tr}
   (\rho(L,\ell) )^\alpha}{1-\alpha},\qquad
    S_{\rm vN}(L,\ell) =\lim_{\alpha\to 1} S_{\rm R} (L,\ell).
\end{gather*}
Here the reduced density matrix $\rho(L,\ell)$
  is obtained from the density matrix of a ground state~$| \Psi \rangle$
by tracing out all spin degrees of freedom
in the environment~B
\begin{gather}
\label{eq:reduced-density-matrix}
\rho(L,\ell) =  \operatorname{Tr}_{\rm B}
\frac{|\Psi \rangle \langle \Psi |}{\langle \Psi | \Psi \rangle }.
\end{gather}
The entanglement spectrum,
i.e.\ the set of the eigenvalues of the
reduced density matrix, determines the entanglement entropies.
For one-dimensional gapless spin chains,
the generic behavior of the entanglement entropies
has been analyzed~\cite{CC} by use of
the conformal f\/ield theory.
The entanglement entropies scale logarithmically
with the size of the subsystem, the prefactor
essentially given by the central charge
of the corresponding conformal f\/ield theory.

On the other hand, gapful chains have been analyzed
by investigating particular models.
One of the most important models is
the Af\/f\/leck--Kennedy--Lieb--Tasaki (AKLT) model~\cite{AKLT}
which was introduced to understand the massive
behavior of integer spin chains~\cite{Haldane1,Haldane2}.
The entanglement entropies of the isotropic AKLT models have been
investigated by examing the exact valence-bond-solid (VBS) ground state~\cite{FKR,KHH,KHK,KKKKT,KX,OT,SK, XKHK}.
For gapped systems which have f\/inite correlation lengths,
the entanglement entropies
saturate at certain values when the size of the subsystems
exceed certain lengths.  The saturated values of
higher rank and higher spin AKLT models are larger than the
spin-1 AKLT model.

Recently Santos et al.\ found surprisingly simple  and useful formula
for calculating the reduced density matrix for  matrix product ground states~\cite{SPKK2, SPKK}.
They applied it to the
AKLT model of spin-1 and general integer spin~$S$
with quantum group symmetry ($q$-AKLT model)
\cite{AM,BMNR, FNW,KSZ1,KSZ2,M,TS1}, and another
 massive Kl\"{u}mper--Schadschneider--Zittartz
 model \cite{KSZ3}
to study anisotropic ef\/fect.

In this article,
we study the entanglement properties of the $q$-AKLT model,
 following the results of~\cite{SPKK2, SPKK}
and giving remarks and additional results.
The more precise def\/inition of the $q$-AKLT model
 on an $L$-site chain with the periodic boundary condition
 is as follows
\begin{gather}
 {\mathcal H}=\sum_{k\in \Z_L}
 \sum_{J=S+1}^{2S}C_J(k,k+1) (\pi_J)_{k,k+1},
\label{hamiltonian}
\end{gather}
where $C_J(k,k+1) > 0$, and $(\pi_J)_{k,k+1}$,
which acts on the $k$-th and $(k+1)$-th sites,
is the $U_q (su(2))$ projection operator
from $V_S \otimes V_S$ to $V_J$, where $V_j$ is the
$(2j+1)$-dimensional highest weight representation of
the quantum group $U_q(su(2))$ \cite{Drinfeld,Jimbo}.
The valence-bond-solid (VBS)
 ground state of this hamiltonian $\mathcal H$
 has a  matrix product form \cite{AM, M},
which generalize the isotropic higher-integer-spin \cite{AAH,FH,TS2}
and spin-1 $q$-deformed AKLT models \cite{BMNR, KSZ1,TS1}.
We check that the entanglement spectra for $\ell=1$
calculated from the formula of the reduced density matrix~\cite{SPKK2, SPKK}
reproduce the one point functions originally derived in~\cite{AM}.
We achieve the f\/inite size corrections of the entanglement entropies
 from the double scaling limit,
 which requires the second order term of the perturbation
 of the entanglement spectrum.
 We exactly calculate the f\/inite size correction term
of the von Neumann entanglement entropy~$S_{\rm vN}(\ell)$.

Besides the entanglement entropies
which characterize the bipartite entanglement,
we also study the geometric entanglement, which is
another kind of measure for entanglement,
see Fig.~\ref{fig:qVBS-EE-GE}.
The geometric entanglement has been proposed as a measure
for multipartite entanglement.
It has been used
to study quantum phase transitions
\cite{Orus1,Orus2,ODV,OT,OW,OWT,SMA, WDMVG,WVG},
and has been measured experimentally recently \cite{ZWL}.
Systems near criticality exhibit logarithmic divergences
as the entanglement entropies. On the other hand,
only a few analytic results are known for gapped systems.
The geometric entanglement def\/ined below can be regarded
as the actual distance between the ground state of the system
and the nearest fully separable state in the Hilbert space.

We divide the $L$-site chain into $N$ parties ($L=N\ell$).
Consider a pure quantum state of
$N$ parties $|\Psi \rangle \in H=\otimes_{i=1}^N H^{[i]}$,
where $H^{[i]}$ is the
space of the $i$th  party.
The entanglement can be quantif\/ied by maximizing the f\/idelity $|\Lambda|$
between the quantum state  $|\Psi \ket$ and  all the possible
separable and normalized states of $N$ parties
$|\Phi \ket=\otimes_{i=1}^N |\phi^{[i]} \ket$, $|\phi^{[i]} \ket
 \in H^{[i]}$,
\begin{gather}
\label{eq:fidelity}
|\Lambda_{\max}|
=\max_{| \Phi \ket } | \Lambda |,
\qquad \Lambda=\frac{\bra \Phi |
\Psi \ket}
{\sqrt{\bra \Psi | \Psi \ket}}.
\end{gather}
The logarithm of $| \Lambda_{\max}|$ is taken
\begin{gather*}
E(\Psi)=
-\Log |\Lambda_{\max}|^2,
\end{gather*}
such that its value becomes zero when $|\Psi \ket$ is separable
or positive otherwise.
The geometric entanglement per block is def\/ined as the
above quantity per party
\begin{gather*}
\mathcal{E}(\ell)= - \lim_{N \to \infty}
  \frac{ E(\Psi) }{N},
\end{gather*}
well def\/ined in the thermodynamic limit.
We evaluate the geometric entanglement for the spin $S$ $q$-deformed
VBS state $|\Psi \ket$. We obtain the expression
of the geometric entanglement
for $\ell \to \infty$ and its f\/inite size corrections
with help of numerical calculations.
For the evaluation of the entanglement entropies and
the geometric entanglement,
 the spectral structure of the transfer matrix
of the $q$-VBS state in the matrix product representation~\cite{AM, M}
will be helpful.

This article is organized as follows.
In Section~\ref{sec:q-VBS},
we brief\/ly review the matrix product representation~\cite{AM, M} of the
VBS ground state of the $q$-AKLT model,
which helps us for evaluating  the entanglement entropies and
the geometric entanglement.
In Section~\ref{sec:EE}, the f\/inite-size correction terms
of the entanglement entropies from the double scaling limit
are calculated
by perturbative analysis.
We emphasize that the double scaling limits
of the entanglement entropies and
 the leading term of the f\/inite-size correction of the entanglement
spectrum have been originally obtained by Santos et al.~\cite{SPKK2}.
But we make Section~\ref{sec:EE} partially overlap their results
so that this article can be self-contained and easy to read.
In Section~\ref{sec:GE}, we
investigate the geometric entanglement with help of numerical calculations.
Section~\ref{section5} is devoted to the summary of this article.

\section[$q$-VBS state]{$\boldsymbol{q}$-VBS state}\label{sec:q-VBS}

In this section, we brief\/ly review the
matrix product representation of the higher-integer-spin
$q$-VBS ground state
and the spectral structure of the transfer matrix of the $q$-AKLT model~\cite{AM, M}.
We use the following notations.
For a real number $c$ we def\/ine
 its $q$ analogue as
\begin{gather*}
    [c]=\frac{q^c-q^{-c}}{q-q^{-1}} .
\end{gather*}
We also def\/ine the $q$-shifted factorial
and the $q$-shifted binomial
for $n\in \mathbb Z_{\ge 0}$ as
\begin{gather*}
    [n]!=
      \begin{cases}
  \displaystyle   \prod_{i=1}^{n} [i],  & n\in {\mathbb N},\\
           1 ,  &  n=0,
      \end{cases}
 \qquad
   \left[ \begin{array}{@{}c@{}}  n \\ k \end{array} \right]
       =\begin{cases}
   \displaystyle   \frac{ [n]! }{ [n]! [n-k]! }, & k=0,\dots,n,\\
           0 ,  &   \rm otherwise .
      \end{cases}
 \end{gather*}

The $q$-VBS state \cite{AM, M}, which is the
exact ground state of the $q$-AKLT model
\eqref{hamiltonian}, is expressed in the following matrix product form
\begin{gather*}
| \Psi \rangle =\operatorname{Tr}
  [g_1 \star g_2 \star \cdots \star g_{L-1} \star g_L],
\end{gather*}
where
$g_k$ is an $(S+1)\times(S+1)$
   vector-valued matrix
acting on the $k$-th site
whose element is given by
\begin{gather*}
g_k(a,b)
 = (-1)^{S-i} q^{(a+b-S)(S+1)/2}
    \sqrt{\left[\begin{array}{@{}c@{}} S \\ a  \end{array} \right]
      \left[ \begin{array}{@{}c@{}} S \\ b \end{array} \right]
    [S-a+b]! [S+a-b]!  }  \;   |S; b-a\rangle_k  \\
\hphantom{g_k(a,b)}{}  =: h_{ab}  |S; b-a  \rangle_k ,
   \qquad  (0\le a,b \le S) .
\end{gather*}
The symbol $\star$ denotes the product
 $A\star B=\left\{ \sum_y|\alpha\rangle_{xy}
 \otimes |\beta\rangle_{yz} \right\}_{xz}$
 for vector-valued matrices
 $A=\left\{ |\alpha\rangle_{xy}\right\}_{xy} $
and  $B= \left\{|\beta\rangle_{xy} \right\}_{xy} $.

We def\/ine $g_k^\dagger$
 by replacing each ket vector in the matrix $g_k$
 by its corresponding bra vector:
\begin{gather*}
   g^\dagger_k(a,b ) = h_{ab} \; {}_k \langle S; b-a  |.
\end{gather*}

Let us set an $(S+1)^2$ dimensional vector space as
\begin{gather*}
  W=\bigoplus_{0 \le a,b \le S} \C|ab  \ketket  ,
\end{gather*}
where
$\{|a b  \ketket \; | \; a,b=0, \dots, S  \}$
 is an orthonormal basis.
We def\/ine an $(S+1)^2 \times (S+1)^2$
   mat\-rix~$G$ acting on the space $W$ as
\begin{gather*}
   G=g^\dagger \otimes g , \qquad
  \brabra a b | G | c d \ketket =
  g^\dagger (a,c) g(b,d)
  = \delta_{a-c,b-d} h_{ac} h_{bd},
\end{gather*}
which plays the role of a transfer matrix.

In \cite{AM}, the spectral  structure of
the $G$ matrix was clarif\/ied, i.e.\
the eigenvalues of $G$ are given as
\begin{gather*}
 \lambda_n =
 (-1)^n (\q{S}! )^2
\left[\begin{array}{@{}c@{}} 2S+1\\S-n
 \end{array}\right]
,\qquad n=0,1,\dots,S,
\end{gather*}
 with the degree of degeneracy $2n+1$,
 and thus
the squared norm of the ground state is given as
\begin{gather}
  \langle\Psi|\Psi\rangle = \operatorname{Tr} G^L
  =  \sum_{0\le n  \le S}
     (2n+1)    \lambda^L_n  . \label{norm}
\end{gather}

The matrix
$G$ has the following block diagonal structure
since $ \brabra ab| G |cd \ketket =0 $
for $ a-b\neq c-d$:
\begin{gather*}
 G= \bigoplus_{-S\le j\le S}G^{ (j) },\qquad G^{ (j) }\in {\rm End} W_j,\qquad
  W_j=
\bigoplus_{i=\max(0,-j)}^{\min(S,S-j)}\C|i,i+j\ketket.
\end{gather*}
The size of each block $G^{ (j) }$ is $(S-|j|+1)\times(S-|j|+1)$.
Each element of $G^{ (j) }$ is
\begin{gather*}
 \brabra a,a+j|G^{ (j) }|c,c+j\ketket
 = (-1)^jq^{(a+c+j-S)(S+1)}   \q{S-a+c}!\q{S+a-c}! \\
 \hphantom{\brabra a,a+j|G^{ (j) }|c,c+j\ketket=}{}   \times \sqrt{
     \left[\begin{array}{@{}c@{}} S \\ a \end{array}\right]
     \left[\begin{array}{@{}c@{}} S \\ a+j \end{array}\right]
     \left[\begin{array}{@{}c@{}} S \\ c \end{array}\right]
    \left[\begin{array}{@{}c@{}} S \\ c+j \end{array}\right]
      }   .
\end{gather*}
We construct intertwiners among the $2S+1$ blocks
 $G^{ (j) }$ $(j=-S, \dots, S)$.
This helps us to construct
eigenvectors of each block from
 another block with a smaller size.

Let us def\/ine a family of linear operators
$\{I_j\}_{ 1\le |j|\le S}$ as
\begin{gather*}
I_j :  \  W_j \to  W_{j-1}  \quad (j>0),\\
\hphantom{I_j :{}}{} \ \brabra a,a+j-1 |I_j| c,c+j \ketket
  =
  \begin{cases}
 \displaystyle  q^{-a}\sqrt{  \frac{\q{a+j}\q{S-a-j+1}}{\q{j}\q{S-j+1}}}, & c=a, \\
\displaystyle  -q^{1-a-j}\sqrt{ \frac{\q{a} \q{S-a+1}}{\q{j}\q{S-j+1}}  }, & c=a-1, \\
  0,                                   & \rm otherwise,
 \end{cases}  \\
\hphantom{I_j :{}}{} \    W_j \to W_{j+1} \quad (j<0), \qquad
\brabra a-j-1,a |I_j| c-j,c \ketket
  =  \brabra a,a-j-1 |I_{-j}| c,c-j \ketket .
\end{gather*}
By direct calculation, one f\/inds
that the matrix $I_j$ enjoys the intertwining relation
$ I_j G^{ (j) } = G^{ (j-1) } I_j$ $(1\le j\le S)$,
$I_j G^{ (j) } = G^{ (j+1) } I_j$  $( -S\le j\le -1)$.
Each block $G^{ (j) }$ has
 a simple (nondegenerated) spectrum
\begin{gather*}
 \operatorname{Spec}  G^{ (j) }= \{ \lambda_\ell \}_{|j| \le  \ell\le S},
\end{gather*}
and the corresponding eigenvectors
are given by
\begin{gather}
    \label{eq:eigen}
|\lambda_n   \ketket_j =
\begin{cases}
\displaystyle   \sum_{0\le i\le S-n}q^{(n+1)i}
   \sqrt{\frac{\q{S-n}! \q{i+n}! \q{S-i}!}{
      [S]! [n]! [S -i-n]! [i]! } }
    |i,i+n\ketket, & n=j\ge 0, \\
\displaystyle  \sum_{0\le i\le S-n}q^{(n+1)i}
   \sqrt{\frac{\q{S-n}! \q{i+n}! \q{S-i}!}{
        [S]! [n]! [S -i-n]! [i]! } }
    |i+n,i\ketket, & -n=j<0, \\
\displaystyle  I_{j+1}I_{j+2}\cdots I_{n} |\lambda_n\ketket_n,
   & 1\le j+1\le n \le S  , \\
 I_{j-1}I_{j-2}\cdots I_{-n} |\lambda_n\ketket_{-n},
 & 1 < -j+1\le n \le S   .
 \end{cases}
\end{gather}
The $\ell$th-power of the $G$ matrix
  is formally expanded as
\begin{gather*}
  G^\ell = \bigoplus_{-S\le j\le S}
  \sum_{  |j| \le  n\le S}
  \frac{\lambda_n^\ell }{ {\ \!}_j\brabra
  \lambda_n|\lambda_n\ketket{\ \!\!}_j  }
  |\lambda_n\ketket{\ \!\!}_j {\ \!}_j\brabra\lambda_n| .
\end{gather*}

\section{Finite size correction of the entanglement entropies}\label{sec:EE}

In this section, we examine the
f\/inite-size correction of the
entanglement entropies
by studying the reduced density matrix.
Recently, the following simple formula for
the reduced density matrix~\eqref{eq:reduced-density-matrix}
 was found~\cite{SPKK}
\begin{gather}
   \rho(L,\ell) =
   \frac{K(L-\ell)K(\ell)}{ \operatorname{Tr} G^L },
\label{eq:Santos-formula}
\end{gather}
where the ``$K$ matrix'' is def\/ined as
\begin{gather*}
K(\ell)  =\mathcal M G^\ell
\end{gather*}
with a linear map $\mathcal M$
\begin{gather}
 \mathcal M \left(
 |ab\ketket\brabra cd | \right) = |ac\ketket\brabra bd | .
\label{eq:map}
\end{gather}
 The reduced density matrix~\eqref{eq:Santos-formula} is an $(S+1)^2 \times (S+1)^2$ matrix,
from which the rank of the density matrix is
equal to or smaller than $(S+1)^2$.
We study the reduced density matrix by combining~\eqref{eq:Santos-formula}
and the spectral structure of the transfer matrix~$G$
reviewed in the last section.

 Here we introduce  some notations and
make some general remarks.
We def\/ine
\begin{gather*}
 K_n   = \sum_{-n\le j\le n}
 \frac{1}{ {\ \!}_j\brabra\lambda_n|\lambda_n\ketket{\ \!\!}_j}
  \mathcal M \left(
 |\lambda_n\ketket{\ \!\!}_j {\ \!}_j\brabra\lambda_n|  \right),
\end{gather*}
so that the $K$ matrix
and the reduced density matrix are written as
\begin{gather}
 K(\ell)  = \sum_{0 \le n\le S}  \lambda_n^\ell K_n,\nonumber\\
 \rho(L,\ell)  =
\frac{1}{\sum\limits_{0\le n\le S} (2n+1) \lambda^L_n}
 \sum_{0 \le n\le S \atop 0 \le n'\le S}
  \lambda_{n'}^{L-\ell} \lambda_n^\ell K_{n'} K_n.
\label{eq:rho-decom}
\end{gather}
One observes that
$K_n $, $K(\ell)$ and $\rho(L,\ell)$
 enjoy the same block diagonal structure as  $G$:
\begin{gather*}
  K_n   = \bigoplus_{-S\le j\le S} K_n^{(j)} ,\qquad
 K(\ell)  = \bigoplus_{-S\le j\le S} K^{(j)} (\ell) ,\qquad
 \rho(L,\ell) =
  \bigoplus_{-S\le j\le S}\rho^{(j)} (L,\ell) ,\\
   K_n^{(j)}   ,     K^{(j)} (\ell),
  \rho^{(j)} (L,\ell)  \in \operatorname{End} W_j,
\end{gather*}
since $\brabra ab| K_n  |cd \ketket =0$
for $ a-b\neq c-d $.
Note that  $\mathcal M$~\eqref{eq:map}
does not always map a matrix acting on a sector $W_j$
to a matrix acting on the same sector.
 The spectrum of  $\rho(L,\ell)$
is, of course, given by the union of
the spectra of $\rho^{(j)}(L,\ell)$'s.
Due to the symmetry
 $ \brabra ab | \rho  (L,\ell) | cd \ketket
 = \brabra ba | \rho  (L,\ell) | dc \ketket $,
we have the degeneracy
\begin{gather*}
\operatorname{Spec}  \rho^{(j)} (L,\ell)
 = \operatorname{Spec}  \rho^{(-j)} (L,\ell)  .
\end{gather*}

\subsection{Double scaling limit}

\begin{figure}[t]
\centering
\includegraphics[width=90mm]{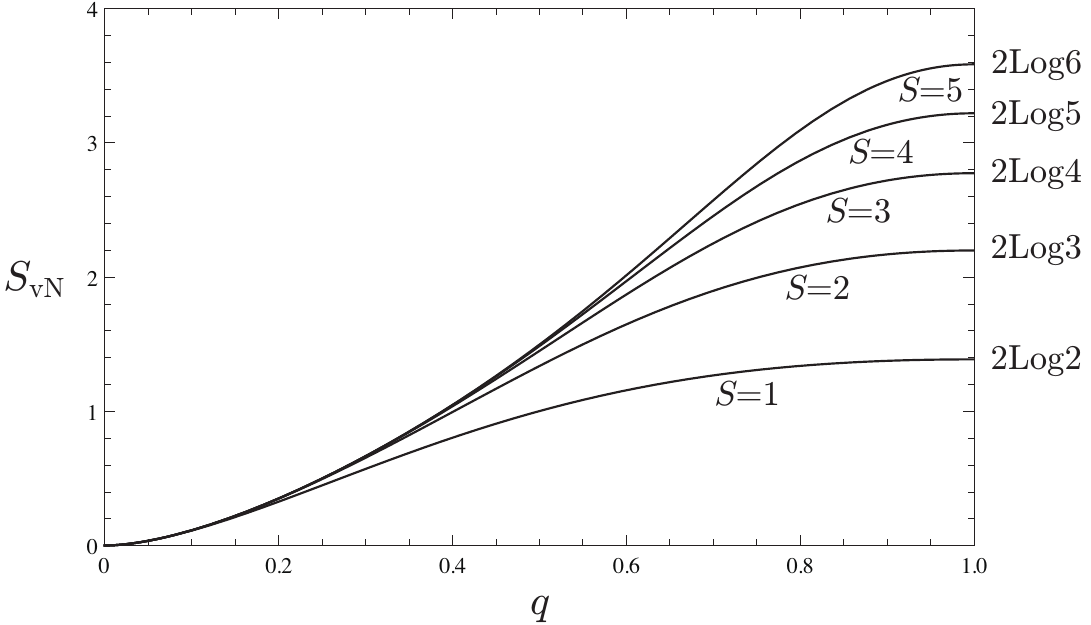}

\caption{The von Neumann entropy
in the double scaling limit $S_{\rm vN}$~\eqref{eq:SvN-DSL}.}
\label{fig:DL}
\end{figure}

We f\/irst review the double scaling limit~\cite{SPKK2, SPKK}
\begin{gather*}
 \rho  =  \lim_{\ell\to\infty} \lim_{L\to\infty}
 \rho(L,\ell) , \qquad
 \rho^{(i)} =   \lim_{\ell\to\infty} \lim_{L\to\infty}
   \rho^{(i)}(L,\ell).
\end{gather*}
Noting the form \eqref{eq:rho-decom}
and $|\lambda_n/\lambda_0| <1$ $(n=1,\dots,S)$,
we f\/ind the reduced density matrix becomes diagonal
\begin{gather}
   \rho = K_0K_0,\qquad
   \brabra ab| \rho | cd \ketket
   =\delta_{ac}\delta_{bd}
   \frac{q^{2(a+b-S)}  }{[S+1]^2}
  \qquad (=: \delta_{ac}\delta_{bd} p_{ab}).
\label{eq:spec-double-limit}
\end{gather}
The entanglement spectrum is, of course,
given by the diagonal elements of $\rho$, i.e.\
$\{  p_{ab}  | a,b =0,1,\dots, S \} $.\footnote{This notation is dif\/ferent from that in \cite{SPKK2}.}
We notice that the degree of
 the degeneracy of the eigenvalue
 $\frac{q^{2k}}{[S+1]^2}$ is $ S-|k|+1$.
For example, the spectrum for $S=2$ is given as
\begin{alignat}{5}
& \operatorname{Spec} \rho^{(2)} :\quad &&
 p_{02} =  \frac{1}{[3]^2}, \qquad & &  && & \nonumber\\
& \operatorname{Spec} \rho^{(1)} :\quad &&
p_{01} =  \frac{1}{q^2 [3]^2},  \qquad &&
 p_{12} =     \frac{q^2}{[3]^2},  \quad && &  \nonumber\\
& \operatorname{Spec} \rho^{(0)} :\quad & &
p_{00} =  \frac{1}{q^4 [3]^2},  \qquad &&
p_{11} =   \frac{1}{[3]^2} , \qquad &&
p_{22} =  \frac{q^4}{[3]^2}, & \nonumber\\
& \operatorname{Spec} \rho^{(-1)} :\quad  &&
p_{10} =  \frac{1}{q^2 [3]^2}, \qquad &&
p_{21} =   \frac{q^2}{[3]^2} , \quad  && &\nonumber \\
& \operatorname{Spec} \rho^{(-2)} :\quad & &
p_{20} =  \frac{1}{[3]^2} . \qquad && && & \label{eq:Spec2-DL}
\end{alignat}
One can calculate
\begin{gather}
\label{eq:P=}
 P := \operatorname{Tr}  \rho^\alpha
=  \sum_{ 0\le a\le S \atop  0\le b \le S  }
  p_{ab}^\alpha
= \left(\frac{[\alpha(S+1)]}{[\alpha][S+1]^\alpha} \right)^2.
\end{gather}
Then we achieve
the entanglement entropies in the double scaling limit \cite{SPKK2, SPKK}
\begin{gather}
S_{\rm R}  = \frac{1}{1-\alpha} \Log P
= \frac{2}{1-\alpha}  \Log
\frac{[\alpha(S+1)]}{[\alpha][S+1]^\alpha}, \nonumber\\
S_{\rm vN}  = 2 \Log q \left(
 \frac{q+q^{-1}}{q-q^{-1}}
-(S+1)\frac{q^{S+1}+q^{-(S+1)} }{ q^{S+1}-q^{-(S+1)} }
\right) + 2\Log [S+1]  ,
\label{eq:SvN-DSL}
\end{gather}
see Fig.~\ref{fig:DL} for the von Neumann entropy
in the double scaling limit.
In particular, when $q=1$,
the spectrum is totally degenerated
\begin{gather}
\label{eq:tot-degen}
p_{ab} = \frac{1}{(S+1)^2},
\end{gather}
and the entropies become
\begin{gather*}
   S_{\rm R}  = S_{\rm vN} = 2\Log(S+1),
\end{gather*}
which agree with the case of the open boundary condition~\cite{KHH,KX,XKHK}.
On the other hand, in the limit $q\to 0$,
only one eigenvalue survives
$p_{00}=1$, $p_{ab}=0$ $(a+b>0)$,
and the entropies become zero.

\subsection{Finite-size correction}

\begin{figure}[t]\centering
\includegraphics[width=75mm]{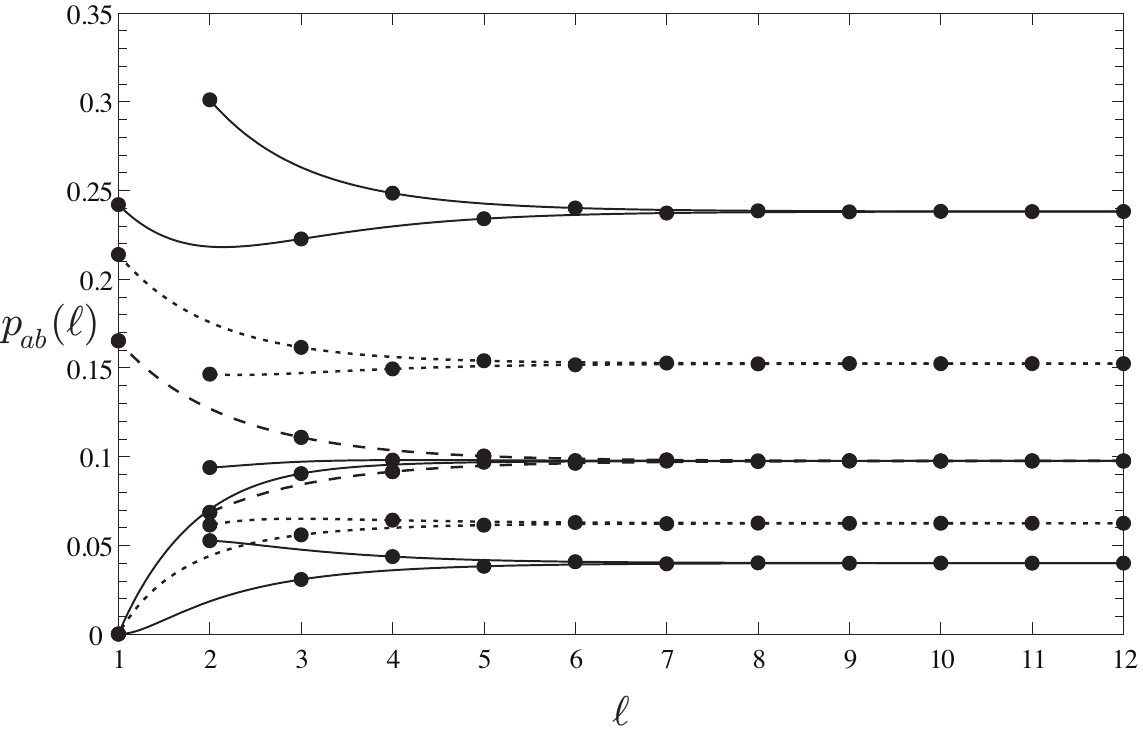}\qquad
\includegraphics[width=75mm]{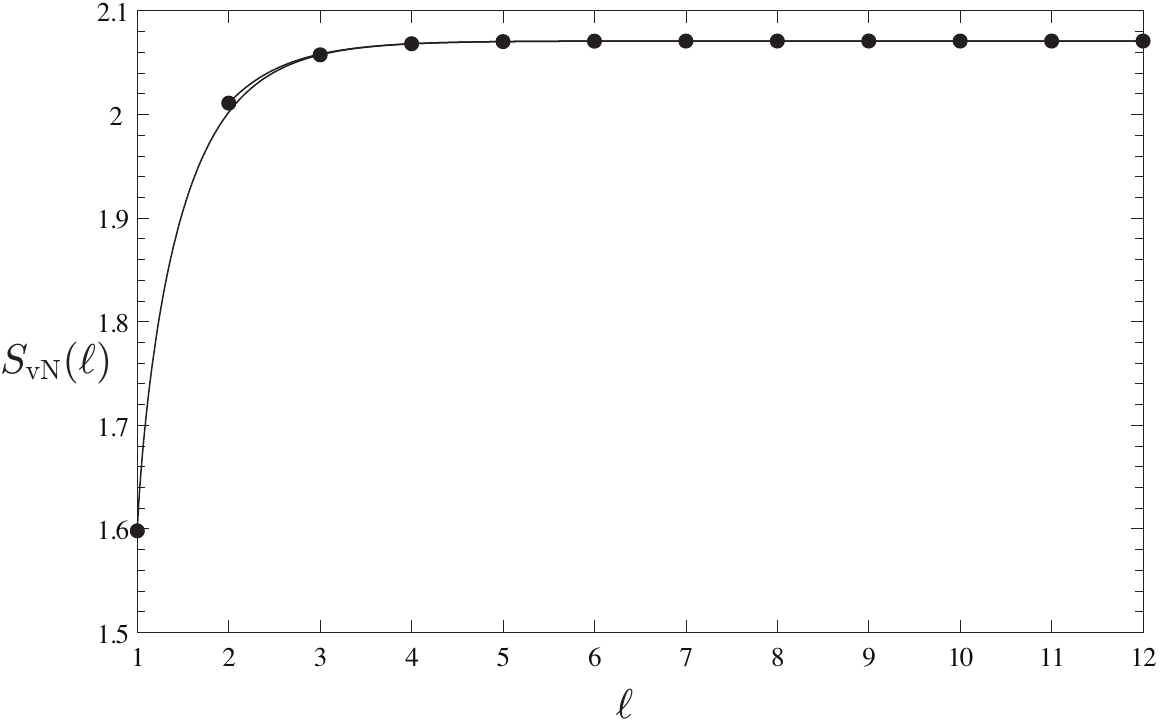}

\caption{The entanglement spectrum of $\rho(\ell)$
(left) and the von Neumann entropy (right)
 for $S=2$ with $q=4/5$.
The lines are drawn for $\ell \in \mathbb R$
 with replacement
 $\kappa_n^\ell \to |\kappa_n|^\ell$
 and  $\kappa_n^\ell \to \kappa_n|\kappa_n|^{\ell-1}$
 in~\eqref{eq:rho-ell-infty} or~\eqref{eq:rho-ell-infty-j}.
In the left f\/igure,
the dashed, dotted and solid lines
correspond to $\operatorname{Spec}\rho^{(\pm 2)}(\ell)$,
 $\operatorname{Spec}\rho^{(\pm 1)}(\ell)$ and
  $\operatorname{Spec}\rho^{(0)}(\ell)$, respectively.}
\label{fig:45}
\end{figure}

We examine the f\/inite-size correction of the entanglement entropies.
We f\/irst take the limit $L\to\infty$
\begin{gather}
\label{eq:rho-ell-infty}
\rho(\ell) :=  \lim_{L\to \infty} \rho(L,\ell) =
 K_0 \sum_{0\le n\le S} K_n \kappa_n^\ell, \\
 \rho^{(j)}(\ell):=
 \lim_{L\to \infty} \rho^{(j)}(L,\ell)  =
 K^{(j)}_0 \sum_{0\le n\le S} K^{(j)}_n
  \kappa_n^\ell,
\label{eq:rho-ell-infty-j}
\end{gather}
with $\kappa_n  = \frac{\lambda_n}{\lambda_0}$,
and then consider the case $\ell=1$
and the behavior of the entropies
for $\ell \to \infty$.
Fig.~\ref{fig:45} provides plots of
the spectrum $\operatorname{Spec}  \rho(\ell) $
of the reduced density matrix~\eqref{eq:rho-ell-infty},
 i.e.\ the union of the spectra
 $\operatorname{Spec} \rho^{(j)}(\ell) $'s
 of \eqref{eq:rho-ell-infty-j},
and the von Neumann entropy
for $S=2$ with $q=4/5$.

For $\ell=1$ and $L\to\infty$
the reduced density matrix becomes
\begin{gather*}
 \rho(1) =   K_0   (\mathcal M G )  /   \lambda_0 .
\end{gather*}
The eigenvalues of $\rho(1)$ become zero
except $2S+1$ ones, which is pointed out
for $S=1$ and~2  in \cite{SPKK2, SPKK}
\begin{gather}
  p_{ab}(1) = \begin{cases}
\displaystyle \delta_{a+b,k}
\frac{[S+|k|]![S-|k|]!}{[2S+1]!}
\sum_{c=0}^{S-|k|} q^{(S+2)(2c+|k|-S)}
\left[\begin{array}{@{}c@{}} S \\ c \end{array}\right]
     \left[\begin{array}{@{}c@{}} S \\ c+|k| \end{array}\right],
&    a\times b=0,  \\  0, & {\rm otherwise} .
  \end{cases}
\label{onesite}
\end{gather}
For example, the spectrum
$\{ p_{ab}(1) \}_{ab}$
of  $\rho(1)$ for $S=2$ is given as
\begin{alignat*}{5}
&\operatorname{Spec} \rho^{(2)}(1): \quad &&
 p_{02} (1) =\frac{1}{[5]}, \qquad && && & \\
&\operatorname{Spec} \rho^{(1)}(1): \quad && p_{01} (1) =\frac{1+q^8}{q^2 (1+q^4)[5]}, \qquad &&
p_{12} (1) =0, &&  &  \\
&\operatorname{Spec} \rho^{(0)}(1):\quad  &&
p_{00} (1)=
  \frac{1-q^2+2 q^6-q^{10}+q^{12}}{q^4(1+q^4 )[5]},\qquad && p_{11}(1)=0, \qquad && p_{22}(1)=0, & \\
&\operatorname{Spec} \rho^{(-1)}(1): \quad &&
p_{10}(1)=  \frac{1+q^8}{q^2 \left(1+q^4\right)[5]}, \qquad &&
p_{21}(1)= 0, \qquad &&   & \\
&\operatorname{Spec} \rho^{(-2)}(1): \quad && p_{20}(1)=\frac{1}{[5]}. \qquad && && &
\end{alignat*}

\begin{figure}[t]\centering
\includegraphics[width=90mm]{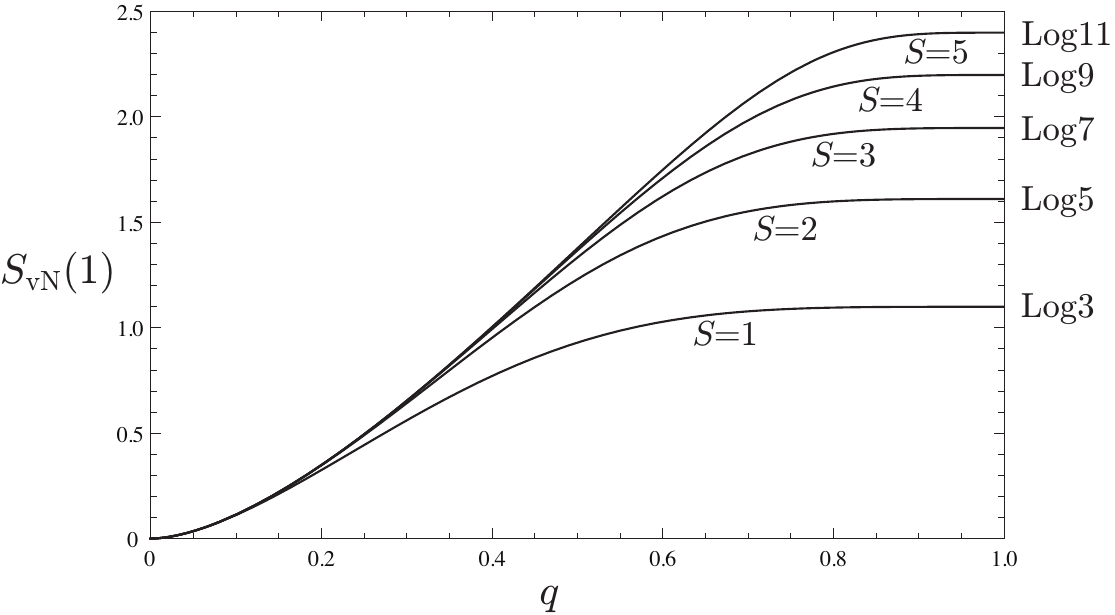}

\caption{The von Neumann entropy
of the one-site subsystem
 $S_{\rm vN} (1)$.}
\label{fig:ell1}
\end{figure}

Let us show \eqref{onesite}. We consider the
submatrix $\rho^{(k)}(1)$, $k \ge 0$. The case for $k<0$ is similar.
By direct calculation, we f\/ind
the matrix elements of $(S-k+1) \times (S-k+1)$ submatrix
$\rho^{(k)}(1)$ are given by
\begin{gather}
 \brabra c,c+k|\rho^{ (k) }(1)|c+j,c+j+k\ketket
 = (-1)^jq^{2c+k-S+(2c+j+k-S)(S+1)}   \frac{\q{S+k}!\q{S-k}!}{[2S+1]!} \nonumber\\
\hphantom{\brabra c,c+k|\rho^{ (k) }(1)|c+j,c+j+k\ketket=}{}
 \times \sqrt{
     \left[\begin{array}{@{}c@{}} S \\ c \end{array}\right]
     \left[\begin{array}{@{}c@{}} S \\ c+j \end{array}\right]
     \left[\begin{array}{@{}c@{}} S \\ c+k \end{array}\right]
     \left[\begin{array}{@{}c@{}} S \\ c+j+k \end{array}\right]
      }  .
\label{onesiteelement}
\end{gather}
The rank of $\rho^{(k)}(1)$ is 1, since
the element
of $\rho^{(k)}(1)$ \eqref{onesiteelement} has a form $A_c \times B_{c+j}$.
Thus, only one eigenvalue of~$\rho^{(k)}(1)$ is nonzero,
which is given by
\begin{gather}
\operatorname{Tr}_{(k)} \rho^{(k)}(1) =\sum_{c=0}^{S-k}
\brabra c,c+k|\rho^{ (k) }(1)|c,c+k\ketket \nonumber \\
\hphantom{\operatorname{Tr}_{(k)} \rho^{(k)}(1)}{}
=\frac{[S+k]![S-k]!}{[2S+1]!}
\sum_{c=0}^{S-k} q^{(S+2)(2c+k-S)}
\left[\begin{array}{@{}c@{}} S \\ c \end{array}\right]
     \left[\begin{array}{@{}c@{}} S \\ c+k \end{array}\right],
\label{onepoint}
\end{gather}
from the fact that the other eigenvalues are all~0.
The expression \eqref{onepoint} is actually
identical to the one point functions
 derived in~\cite{AM}.
In particular, when $q=1$,
the non-zero eigenvalues are degenerated as
$  p_{ab}(1) = \frac{1}{2S+1}$ $(a\times b=0) $,
and we have
$   S_{\rm R} (1)  =  S_{\rm vN} (1)   =  \Log (2S+1) $
\cite{SPKK2}.
One observes the monotonicity of
the von Neumann entropy $S_{\rm vN} (1)$
 while $0<q<1$, see Fig.~\ref{fig:ell1}.

We turn to the behavior of entropies
for $\ell\to\infty$.
Noting again the form \eqref{eq:rho-decom}
and $|\kappa_n| <1$ $(n=1,\dots,S)$,
we f\/ind
\begin{gather*}
 \rho (\ell )
 = \rho  + \kappa_1^\ell K_0 K_1  +  O \big(  \kappa_2^\ell \big),
 \qquad  \ell\to\infty    .
\end{gather*}

We denote the eigenvalue of $\rho (\ell) $
by $ p_{ab} (\ell) $ corresponding to
 $ p_{ab} $ \eqref{eq:spec-double-limit}
 when $\ell\to\infty$.
Since the density matrix $\rho$
in the double scaling limit is a diagonal matrix,
it is not dif\/f\/icult to perform perturbative calculation.
Noting $ |\kappa_1 |^2  > | \kappa_2 |> | \kappa_3 |
>\cdots $, we f\/ind
\begin{gather}
\label{eq:spec-pert}
   p_{ab}(\ell)   =   p_{ab}
 +  r_{ab} \kappa_1^\ell  +  t_{ab} \kappa_1^{2\ell}
 + o\big(\kappa_1^{2\ell}\big), \qquad  \ell\to\infty , \\
  r_{ab}  = \brabra ab | K_0K_1 | ab \ketket ,\qquad
t_{ab} =
 \sum_{(c,d)\neq (a,b) \atop 0\le c,d\le S}
  \frac{\brabra cd | K_0K_1 | ab \ketket
   \brabra ab | K_0K_1 | cd \ketket}{p_{ab}-p_{cd}} .\nonumber
 \end{gather}
Inserting  \eqref{eq:eigen}  into
 $r_{ab}$ and $t_{ab}$ def\/ined above,
 we have
\begin{gather}
\label{eq:spec-pert-r}
r_{ab} =\frac{q^{ a+b-S }}{[S+1] }
 \frac{q^{a+b-3S}[3](1-q^{2a}-q^{2a+2}+q^{2S+2})(1-q^{2b}-q^{2b+2}+q^{2S+2})}{(q^2-1)^2 [S][S+1][S+2]},  \\
t_{ab}  =
\left(\frac{q^{2(a+b-S)}[3] [2] }{[S+2][S+1][S]}\right)^2\nonumber\\
\hphantom{t_{ab}  =}{}
\times\frac{q^{-4}[S-a+1][a][S-b+1][b]-q^4[S-a][a+1][S-b][b+1]}{q^2-q^{-2} }.\label{eq:spec-pert-t}
\end{gather}
The f\/irst-order term \eqref{eq:spec-pert-r}
has been originally  obtained in~\cite{SPKK2}
(see equation~(59) of~\cite{SPKK2} by changing the indices
$\mu=S/2-a$, $\nu=S/2-b$ and redef\/ining $q \to q^{1/2}$),
where the characteristic length
is given by $\xi=\frac{1}{\Log ( [S+2]/[S] ) }$.
We also calculated the second-order term~\eqref{eq:spec-pert-t} which
is needed for seeing the f\/inite-size correction of
the von Neumann entropy.

For example, the spectrum $\{p_{ab} (\ell)\}_{ab}$
($\ell\to\infty$)
for $S=2$ (which is shifted from \eqref{eq:Spec2-DL}
as \eqref{eq:spec-pert})
 is given as
\begin{alignat*}{3}
& \operatorname{Spec} \rho^{(2)}(\ell): \quad &&
 \frac{1}{[3]^2}-\frac{[2]}{[3][4]}
 \kappa_1^\ell,
  &    \\
& \operatorname{Spec} \rho^{(1)}(\ell): \quad &&
 \frac{1}{q^2 [3]^2} +\frac{ (1-q^2)[2]}{q^2 [3] [4]}\kappa_1^\ell
 +\frac{q^2 [2]^2}{ (1-q^4 ) [4]^2}
  \kappa_1^{2\ell},
  & \\
  &&& \frac{q^2}{[3]^2}-\frac{ (1-q^2) [2]}{[3] [4]}\kappa_1^\ell
  -\frac{q^2[2]^2}{ (1-q^4 ) [4]^2}
  \kappa_1^{2\ell} , &   \\
& \operatorname{Spec} \rho^{(0)}(\ell): \quad &&
 \frac{1}{q^4 [3]^2}+\frac{[2]}{q^2 [3] [4]}\kappa_1^\ell
 +\frac{ [2]^2}{q^2  (1-q^4 ) [4]^2}
 \kappa_1^{2\ell} ,  & \\
&&&  \frac{1}{[3]^2}+\frac{(1-q^2)^2  [2]}{q^2 [3] [4]}\kappa_1^\ell
  -\frac{[2]}{[4]}\kappa_1^{2\ell} , \qquad  \frac{q^4}{[3]^2}+\frac{q^2[2]}{[3] [4]} \kappa_1^\ell
  -\frac{q^6 [2]^2}{(1-q^4 ) [4]^2}
  \kappa_1^{2\ell}  , & \\
& \operatorname{Spec} \rho^{(-1)}(\ell): \quad &&
 \frac{1}{q^2 [3]^2}+\frac{ (1-q^2)   [2]}{q^2 [3] [4]} \kappa_1^\ell
 +\frac{q^2[2]^2}{(1-q^4 )[4]^2}
  \kappa_1^{2\ell} , & \\
  &&&
  \frac{q^2}{[3]^2}-\frac{ (1-q^2)   [2]}{[3] [4]} \kappa_1^\ell
  -\frac{q^2 [2]^2}{ (1-q^4 ) [4]^2} \kappa_1^{2\ell} ,  &  \\
& \operatorname{Spec}\rho^{(-2)}(\ell): \quad &&
 \frac{1}{[3]^2}-\frac{[2]}{[3] [4]}
  \kappa_1^\ell,    &
\end{alignat*}
where we omit the symbol $+o\big(\kappa_1^{2\ell} \big)$.

The R\'enyi entropy
is expressed by $p_{ab}$, $r_{ab}$ and $t_{ab}$
up to the order of $\kappa_1^{2\ell}$ as
\begin{gather}
S_{\rm R} (\ell)  =
\frac{1}{1-\alpha} \Log\operatorname{Tr}
 \left(  \rho(\ell)\right)^\alpha =
\frac{1}{1-\alpha} \Log
 \sum_{ 0\le a\le S \atop  0\le b \le S  }
 \left(p_{ab}(\ell)\right)^\alpha
\nonumber \\
 \hphantom{S_{\rm R} (\ell)}{}
   =  \frac{1}{1-\alpha} \Log
   \big\{P \big( 1 + R \kappa^\ell_1 + T \kappa_1^{2\ell} \big)
    + o\big(\kappa_1^{2\ell} \big) \big\} \nonumber \\
\hphantom{S_{\rm R} (\ell)}{}
    =  S_{\rm R}
+ \frac{R}{1-\alpha}  \kappa^{\ell}_1
   +\frac{1}{1-\alpha} \left(T-\frac{R^2}{2}\right)
    \kappa_1^{2\ell}
   + o\big(\kappa_1^{2\ell}\big),  \qquad \ell\to\infty,
\label{eq:SR-ell}
\end{gather}
where
\begin{gather*}
  R =   \frac{\alpha}{P}
   \sum_{ 0\le a\le S \atop  0\le b \le S  }
        p_{ab}^{\alpha-1}  r_{ab},\qquad
  T =  \frac{\alpha}{P}
   \sum_{ 0\le a\le S \atop  0\le b \le S  }
 \left( p_{ab}^{\alpha-1} t_{ab} + \frac{\alpha-1}{2}
 p_{ab}^{\alpha-2} r_{ab}^2  \right),
\end{gather*}
with $P$ def\/ined by \eqref{eq:P=}.

By tedious but straightforward calculation, one f\/inds
\begin{gather*}
R=  \frac{\alpha[3]}{[S][S+2]}
\left(\frac{[S+2][\alpha S]-[\alpha(S+2)][S]}
{(q-q^{-1})[\alpha(S+1)][\alpha+1]} \right)^2, \\
T=
\alpha \left( \frac{[2][3]}{(q-q^{-1})^2 [S][S+2]} \right)^2
 \left[
\frac{[2(\alpha-1)]}{[2]} \left(
\frac{ ( [(S + 2) (\alpha + 1)] [S] -
        [S + 2] [S (\alpha + 1)] )}{[\alpha+2]
       [\alpha+1] [\alpha (S + 1)] }\right)^2\right.
  \nonumber \\
\left.
\hphantom{T=}{}  +\frac{\alpha-1}{2}\left( \frac{[2(S+1)]^2}{[2][S+1]^2}
-\frac{ 2 [2(S+1)] [\alpha] [(\alpha+1)(S+1)] }{
 [S+1] [\alpha(S+1)] [\alpha+1]  } +
\frac{ [2] [\alpha] [ (\alpha+2)(S+1)] }{
  [(S+1) \alpha] [\alpha+2]  } \right)^2 \right].
\end{gather*}
Then we f\/ind
\begin{gather}
\label{eq:SvN-ell}
   S_{\rm vN} (\ell)   = S_{\rm vN}
   + \left( \frac{4q^2 \Log q}{1-q^4}
     - \frac{1}{2} \right) \kappa_1^{2\ell}
      + o\big(\kappa_1^{2\ell}\big),
     \qquad  \ell\to\infty ,
\end{gather}
where the coef\/f\/icient of $\kappa_1^{\ell}$ vanishes.
 Since the leading order term is  $\kappa_1^{2\ell}$,
 the characteristic length is~$2\xi$.
We f\/ind  the coef\/f\/icient of~$\kappa_1^{2\ell}$
  depends on the anisotropy parameter~$q$
 but is independent of the spin value~$S$.

As discussed in \cite{SPKK2},
the perturbation fails for the isotropic case
due to the degeneracy~\eqref{eq:tot-degen}, but
the entanglement spectrum can be written by
linear combinations of $\kappa_n$'s and
has the same spectral structure for the transfer matrix~$G$.
For example, for $S=2$, we have
\begin{alignat*}{5}
&\operatorname{Spec} \rho^{(2)} (\ell) : \quad &&
 \frac{1}{3}  \left(\frac{1}{3}-\frac{\kappa_1^\ell}{2}
 +\frac{\kappa_2^\ell}{6}\right), \qquad && && & \\
&\operatorname{Spec} \rho^{(1)} (\ell) : \quad &&
  \frac{1}{3} \left(\frac{1}{3}-\frac{\kappa_1^\ell}{2}
 +\frac{\kappa_2^\ell}{6}\right),  \qquad &&
  \frac{1}{3} \left(\frac{1}{3}+\frac{\kappa_1^\ell}{2}
 -\frac{5\kappa_2^\ell}{6}\right), \qquad &&  & \\
&\operatorname{Spec} \rho^{(0)} (\ell) : \quad &&
  \frac{1}{3} \left(\frac{1}{3}-\frac{\kappa_1^\ell}{2}
 +\frac{\kappa_2^\ell}{6}\right), \qquad & &
  \frac{1}{3} \left(\frac{1}{3}+\frac{\kappa_1^\ell}{2}
 -\frac{5\kappa_2^\ell}{6}\right),  \qquad &&   \frac{1}{3} \left(\frac{1}{3}+ \kappa_1^\ell
  +\frac{5\kappa_2^\ell}{3}\right), & \\
&\operatorname{Spec} \rho^{(-1)} (\ell) : \quad &&
  \frac{1}{3} \left(\frac{1}{3}-\frac{\kappa_1^\ell}{2}
 +\frac{\kappa_2^\ell}{6}\right),  \qquad &&
  \frac{1}{3} \left(\frac{1}{3}+\frac{\kappa_1^\ell}{2}
 -\frac{5\kappa_2^\ell}{6}\right),  && &   \\
&\operatorname{Spec} \rho^{(-2)} (\ell) : \quad &&
  \frac{1}{3} \left(\frac{1}{3}-\frac{\kappa_1^\ell}{2}
 +\frac{\kappa_2^\ell}{6}\right) , \qquad && && &
\end{alignat*}
where no higher order term is needed.
In \cite{SPKK2} the f\/inite-size corrections of the entanglement entropies
for $q=1$ were calculated as
\begin{gather*}
  S_{\rm R} (\ell)  =
  2\Log (S+1)    - \frac{3}{2} \alpha \kappa_1^{2\ell}
   +o\big(\kappa_1^{2\ell}\big),\qquad  \ell\to\infty , \\
  S_{\rm vN} (\ell)  =
  2\Log (S+1)    - \frac{3}{2} \kappa_1^{2\ell}
   +o\big(\kappa_1^{2\ell}\big),\qquad  \ell\to\infty  ,
\end{gather*}
which agree with the limits $q\to 1$ of
\eqref{eq:SR-ell} and \eqref{eq:SvN-ell}.

\section{Geometric entanglement}\label{sec:GE}

In this section, we evaluate the geometric
entanglement, which is another kind of measure of
entanglement.
We divide  the chain into $N$ parties ($L=N\ell$),
and each of the $N$ parties to be contiguous blocks of~$\ell$ spins~$S$.
When $N$ is large enough,
the following expression for the f\/idelity
 $|\Lambda_{\max}|$ \eqref{eq:fidelity}
has been shown for $PT$-symmetric matrix product ground states
$| \Psi \ket$ in~\cite{Orus1,Orus2,OW}
\begin{gather}
|\Lambda_{\max}|^2 =\lim_{N \to \infty}
\frac{|d|^{2N}}{\bra \Psi |\Psi \ket},
\label{eq:fidelityexp}
\end{gather}
where $|d|^2$ is the quantity
\begin{gather}\label{maximization}
|d|^2 =\max_{\{x_i\}:\,\sum\limits_{i=0}^S |x_i|^2=1}
 \bra \operatorname{Aux} |G^\ell| \operatorname{Aux} \ket,\qquad
|\operatorname{Aux} \ket =\sum_{0\le a\le S\atop 0\le b\le S} x_a x_b^*|ab \ketket .
\end{gather}
Performing the maximization \eqref{maximization},
one obtains the f\/idelity $|\Lambda_{\max}|$
which f\/inally leads to the analytic expression for the
geometric entanglement $\mathcal{E}(\ell)$
\begin{gather*}
\mathcal{E}(\ell)  = -\lim_{N \to \infty}
    \frac{\Log |\Lambda_{\max}|^2}{N} .
\end{gather*}
For convenience we set $x_i=r_ie^{ \sqrt{-1}\theta_i }$
($r_i\ge0$, $\theta_i\in\mathbb R$), and write
$x^\blt_i=r^\blt_ie^{\sqrt{-1}\theta^\blt_i}$
if the setting  $\{x_i=x^\blt_i\}$ maximizes
$\langle\operatorname{Aux} |G^\ell| \operatorname{Aux}\rangle$.

\begin{figure}[t]\centering
 \includegraphics[width=75mm]{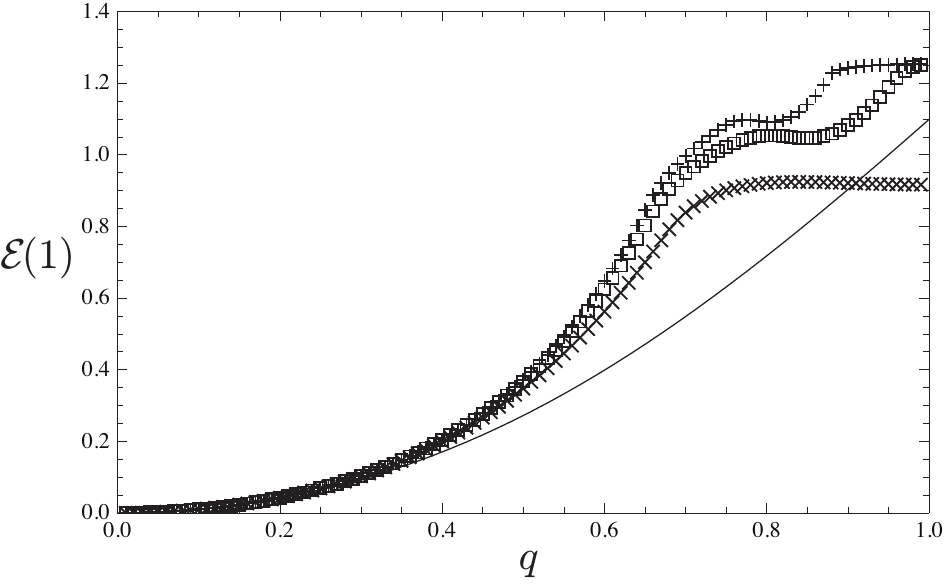}\qquad
 \includegraphics[width=75mm]{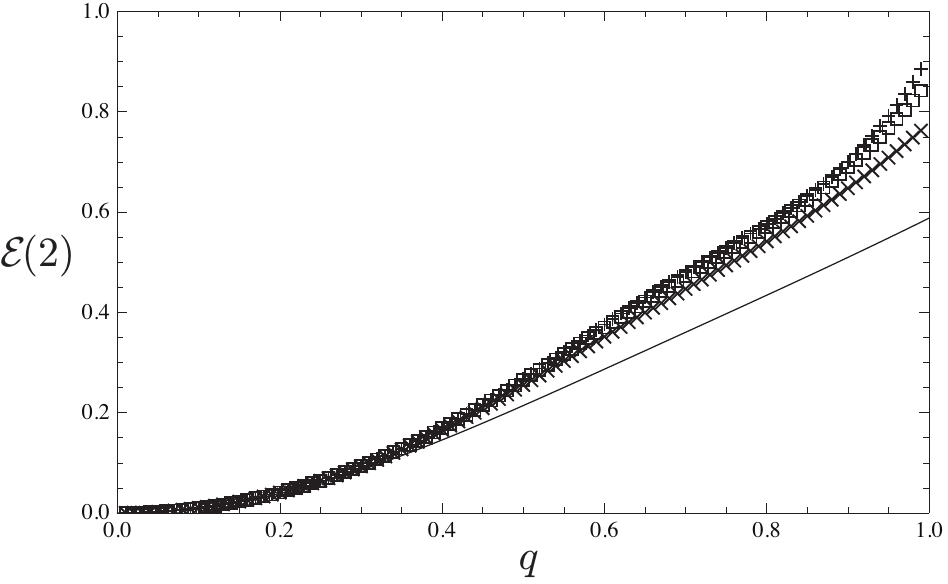}

\caption{The geometric entanglements for $\ell=1$ (left) and 2 (right).
In each graph, the line  corresponds to $S=1$ \eqref{eq:GE-S=1},
and the markers $\times$ ($S=2$),
  $\square$ ($S=3$) and  $+$ ($S=4$)
are plotted based on numerical calculations.}\label{fig:GEonetwo}
\end{figure}

\subsection{Spin-1}

We calculate the geometric entanglement for $S=1$.
By direct calculation, we have
\begin{gather*}
  \bra \operatorname{Aux} |G^\ell| \operatorname{Aux} \ket
=  \left( \frac{[3]^\ell}{q[2]}+\frac{(-1)^\ell}{q^{-1}[2]} \right)|x_0|^4
+\left( \frac{2[3]^\ell}{[2]}+2(-1)^\ell -\frac{2(-1)^\ell}{[2]} \right)|x_0|^2|x_1|^2 \nonumber \\
\hphantom{\bra \operatorname{Aux} |G^\ell| \operatorname{Aux} \ket  =}{}
+\left( \frac{q[3]^\ell}{[2]}+\frac{q^{-1}(-1)^\ell}{[2]} \right)|x_1|^4.
\end{gather*}
Inserting $r_1^2=1-r_0^2$, we get
\begin{gather*}
\bra \operatorname{Aux} |G^\ell| \operatorname{Aux} \ket
= \frac{1-q}{1+q^2}\big([3]^\ell-(-1)^\ell\big)
\left(  (1-q) r_0^4 + 2q r_0^2 \right)
+\frac{q^2[3]^\ell}{1+q^2}+\frac{ (-1)^\ell}{1+q^2},
\end{gather*}
where $\theta_i$'s do not appear.
Thus we f\/ind
\begin{alignat}{3}
&0<q<1: \quad && r_0^\blt = 1, \qquad
          |d|^2=\frac{[3]^\ell}{1+q^2}+\frac{q^2(-1)^\ell}{1+q^2}, &\nonumber\\
 &    q=1:\quad &&
         |d|^2=  \frac{  3^\ell}{2}+\frac{ (-1)^\ell}{2}   ,& \nonumber\\
&    q>1: \quad &&  r_0^\blt = 0, \qquad
          |d|^2=\frac{q^2[3]^\ell}{1+q^2}+\frac{ (-1)^\ell}{1+q^2}, &\label{eq:GE-S=1}
\end{alignat}
where $|d|^2=\bra \operatorname{Aux} |G^\ell| \operatorname{Aux} \ket$
is independent of $\{x_i\}$ at  the isotropic point  $q=1$ \cite{Orus1},
and the choice of $r^{\blt}_0$ changes discontinuously
at this point.
(We will see that
this kind of ``degeneracy''  occurs for the higher spin case.)
Inserting these forms and $\bra \Psi | \Psi \ket=[3]^{L}+3(-1)^{L}$
into \eqref{eq:fidelityexp}, we f\/inally achieve the geometric entanglement
\begin{alignat*}{3}
&  0<q<1: \quad &&
  \mathcal{E}(\ell)= \Log\big(1+q^2\big)-\Log \big(1+q^2  (- [3])^{-\ell} \big), & \\
&  q=1: \quad &&
  \mathcal{E}(\ell)= \Log2 - \Log \big(1+ (- [3])^{-\ell} \big) , & \\
&  q>1: \quad &&
  \mathcal{E}(\ell)= \Log\big(1+q^2\big)-\Log \big( q^2 + (- [3])^{-\ell} \big),
\end{alignat*}
which generalizes~\cite{Orus1}.
The entanglement entropy takes its maximum at the isotropic point,
 decreases with the decrease of the anisotropy parameter $q$
and f\/inally becomes $\mathcal{E}(\ell)=0$ at $q=0$,
see Fig.~\ref{fig:GEonetwo}.
This behavior of the geometric entanglement is similar to the
entanglement entropies.
In the limit $\ell\to\infty$, we have
\begin{gather*}
  \mathcal{E} = \Log \big(1+q^2\big).
\end{gather*}

\subsection{Spin-2}

Let us consider f\/irst the isotropic case,
where we have
\begin{gather*}
 \bra \Aux  |G^\ell| \Aux  \ket  \\
 \qquad{} =   \frac{1}{3} \lambda_0^\ell +\frac{1}{2} \lambda_1^\ell
 +\frac{1}{6} \lambda_2^\ell
 - \frac{  (1 - (r_0-r_2)^2 )^2
 - 4 r_0r_1^2r_2(\cos(\theta_0-2\theta_1+\theta_2)+1) }{2}
   \big( \lambda_1^\ell -  \lambda_2^\ell \big)   \\
 \qquad{}  =  \frac{1}{3} \lambda_0^\ell +\frac{1}{2} \lambda_1^\ell
 +\frac{1}{6} \lambda_2^\ell
 - \frac{(r_1^2-2r_0r_2)^2
 -4 r_0r_1^2r_2(\cos(\theta_0-2\theta_1+\theta_2)-1) }{2}
   \big( \lambda_1^\ell -  \lambda_2^\ell\big )
\end{gather*}
with $\lambda_0=40$, $\lambda_1=-20$ and $\lambda_2=4$.
When $\ell$ is odd (resp.\ even),
 $\lambda_1^\ell < \lambda_2^\ell$
(resp. $\lambda_1^\ell > \lambda_2^\ell$).
Using the f\/irst (resp.\ second) form, we f\/ind
\begin{alignat*}{3}
&  \ell \ {\rm odd} :\quad &&
r_0^{\blt  } -  r_2^\blt= 0 ,\qquad
\cos(\theta^\blt_0 - 2 \theta^\blt_1 + \theta^\blt_2) =-1,\qquad
 |d|^2  = \frac{1}{3} \lambda_0^\ell  +\frac{2}{3} \lambda_2^\ell, & \\
& \ell \ {\rm even} :\quad &&
r_1^{\blt 2} - 2 r_0^\blt r_2^\blt= 0 ,\qquad
\cos(\theta^\blt_0 - 2 \theta^\blt_1 + \theta^\blt_2) =1,\qquad
 |d|^2  = \frac{1}{3} \lambda_0^\ell +\frac{1}{2} \lambda_1^\ell +\frac{1}{6} \lambda_2^\ell.&
\end{alignat*}

\begin{figure}[t]\centering
 \includegraphics[width=75mm]{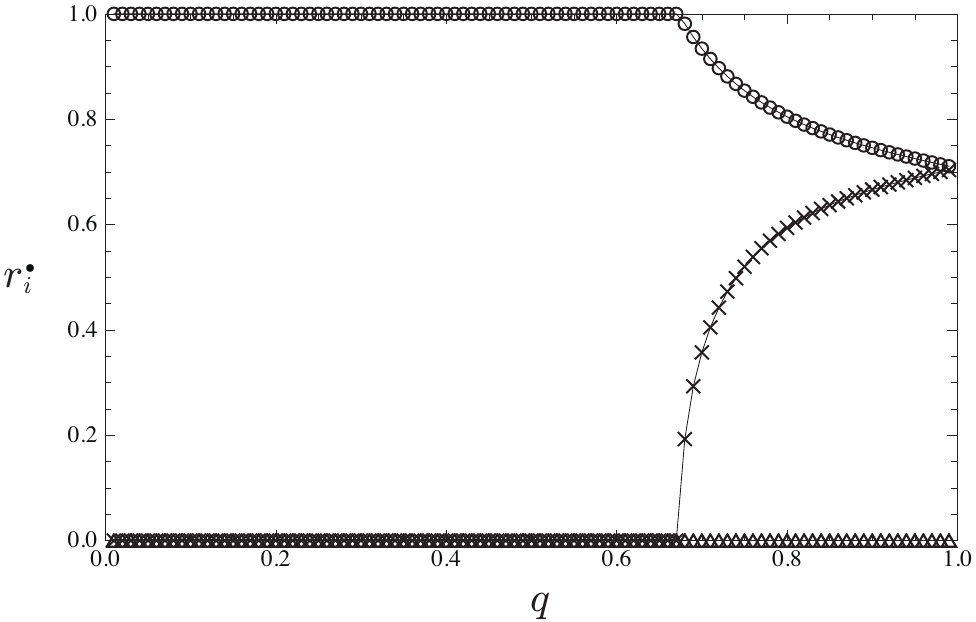}
\qquad
 \includegraphics[width=75mm]{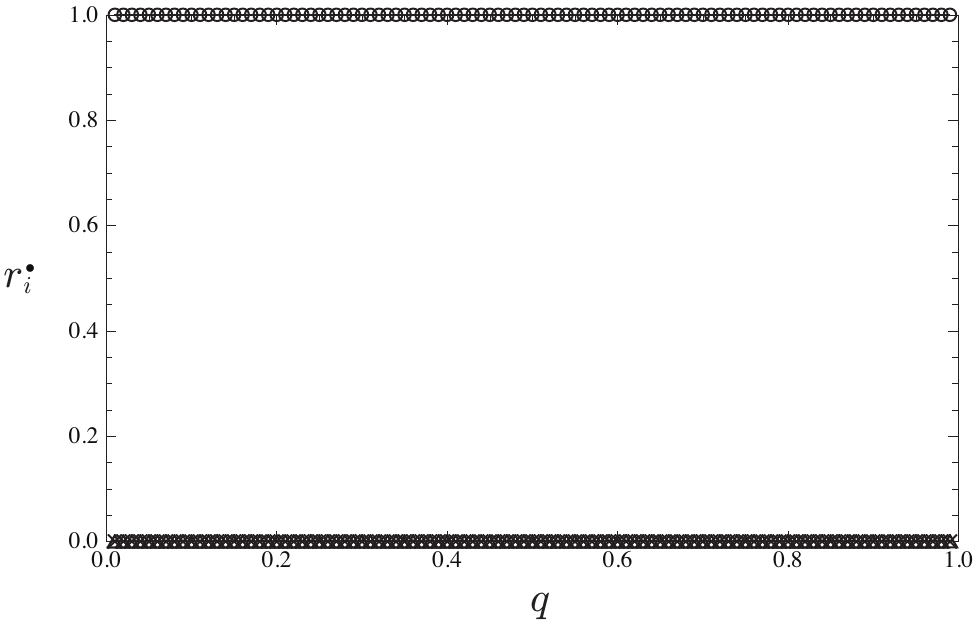}

\caption{Numerical calculations of
the sets $\{r_i^\bullet\}$
which maximize $|d|^2$ for $S=2$ and $\ell=1$ (left)
and 2 (right).
The markers $\bigcirc$, $\triangle$ and $\times$
correspond to $r_0^\bullet$, $r_1^\bullet$ and $r_2^\bullet$, respectively.}\label{fig:S2-ell12}
\end{figure}

In the anisotropic case, thanks to the form
\begin{gather*}
  \langle \Aux| G^\ell| \Aux \rangle
  = \frac{4q^2r_0r_1^2r_2}{1+q^4}\big(\lambda_1^\ell-\lambda_2^\ell\big)
  \cos (\theta_0-2\theta_1+\theta_2)
     + \text{(independent of} \ \{ \theta_i \}),
\end{gather*}
we have
\begin{alignat*}{3}
&  \ell \ {\rm odd} :\quad &&
\cos(\theta^\blt_0 - 2 \theta^\blt_1 + \theta^\blt_2) =-1, & \\
& \ell \ {\rm even} :\quad &&
\cos(\theta^\blt_0 - 2 \theta^\blt_1 + \theta^\blt_2) =1, &
\end{alignat*}
which is the same as for the isotropic case.
We use help of numerical calculations
(see Fig.~\ref{fig:S2-ell12} for $\ell=1$ and 2),  which indicates that
\begin{alignat*}{3}
& \ell \ {\rm  odd}: \quad &&
      r_i^\blt = \delta_{i0} \ {\rm  for} \ q\le\exists \, q^\blt ; \
         r_0^\blt>0, \ r_1^\blt = 0, \ r_2^\blt>0 \   {\rm for} \  q^\blt<q<1 , &\\
& \ell \ {\rm even} :  \quad &&   r_i^\blt = \delta_{i0} \
     \text{always maximizes} \  \langle \Aux| G^\ell| \Aux \rangle . &
\end{alignat*}
One observes that the geometric entanglement with $\ell$ odd
is not completely monotonic while $0<q<1$, see Fig.~\ref{fig:GEonetwo}.
The set $\{r_0^\blt,r_2^\blt\}$ is obtained by
\[
\frac{d}{dr_0} \left(\langle \Aux| G^\ell| \Aux \rangle
\big|_{r_1=0,r_2=\sqrt{1-r_0^2 },
\cos(\theta_0-2\theta_1+\theta_2) =1 }\right)  = 0
\]
in the case where $\ell$ is odd and  $q^\blt<q<1$
\begin{gather}\label{eq:r0}
  r_0^\blt = q
  \sqrt{\frac{-(1-q^2)(1+q^4)\lambda_0^\ell+2[3]\lambda_1^\ell-(1+3q^2+q^4+q^6)\lambda_2^\ell}
  {(1-q^2)^2(1+q^4)\lambda_0^\ell+4q^2[3]\lambda_1^\ell-(1+2q^2+6q^4+2q^6+q^8)\lambda_2^\ell} } .
\end{gather}
The transition point $q^\blt$ is obtained by solving \eqref{eq:r0}~$=1$,
which approaches~1 as $\ell\to\infty$.
The set~$\{r_i^\blt\}$ for $q>1$
 is obtained by replacing $r_0\leftrightarrow r_1$
and $q\to 1/q$.
Under the assumption $r_i^\blt =\delta_{i0}$, we have
\begin{gather*}
   \mathcal{E}(\ell)  = - \Log \frac{ |d|^2 }{\lambda_0^\ell}
   = -  \Log  \langle\Aux| G^\ell |\Aux\rangle |_{r_i=\delta_{i0}}
  \\
\hphantom{\mathcal{E}(\ell)}{}
= \Log(1+q^2+q^4)
  - \Log  \left( 1+ \frac{q^2(1+q^2+q^4)\kappa_1^\ell}{1+q^4}
    + \frac{q^8\kappa_2^\ell}{1+q^4}  \right)  ,
\end{gather*}
for $0<q<1$ and suf\/f\/iciently large $\ell$.

\subsection{Spin-3}

For the isotropic case, we have
\begin{gather*}
  \bra \Aux  |G^\ell| \Aux  \ket
     =  \frac{1}{4} \lambda_0^\ell+ \frac{9}{20} \lambda_1^\ell
  +\frac{1}{4} \lambda_2^\ell + \frac{1}{20}  \lambda_3^\ell
  -\frac{A}{5} ( \lambda_1^\ell -  \lambda_3^\ell ), \\
  A =
  \Big[   (r_1 r_2 - 3 r_0 r_3)^2 + 2\big(r_2^2 - \sqrt{3} r_1 r_3\big)^2 +
 2\big(r_1^2 - \sqrt{3} r_0 r_2\big)^2 \\
 \hphantom{A=}{}
 - 2 r_1 r_2 \big\{
       3 r_0 r_3(\cos(\alpha+\beta)-1)
    +   2 \sqrt{3} r_0 r_1(\cos\alpha -1)
  +2\sqrt{3}r_2 r_3  (\cos\beta-1) \big\}
 \Big]  ,
\end{gather*}
where $\alpha=\theta_0 - 2 \theta_1 + \theta_2$
and $\beta=\theta_1 - 2 \theta_2 + \theta_3$.
When $\ell$ is even, $\lambda_1^\ell > \lambda_3^\ell$.
Thus $  \bra \Aux  |G^\ell| \Aux  \ket$ is maximized by
\begin{gather*}
 r_2^{\blt 2} - \sqrt{3} r^\blt_1 r^\blt_3
= r_1^{\blt 2}- \sqrt{3} r^\blt_0 r^\blt_2=0 ,\qquad
\cos \alpha^\blt =\cos \beta^\blt =1 ,  \\
 |d|^2
 =  \frac{1}{4} \lambda_0^\ell+ \frac{9}{20} \lambda_1^\ell
  +\frac{1}{4} \lambda_2^\ell + \frac{1}{20}  \lambda_3^\ell  .
\end{gather*}
When $\ell$ is odd,
the candidates of $\{\theta_i\}$ that maximize $\bra\Aux|G^\ell|\Aux\ket$
for given $\{r_i\}$ are
\begin{gather}
 \cos\alpha =
   \frac{3r_2^2r_3^2-4r_1^2r_2^2-3r_0^2r_1^2}{4\sqrt{3}r_0r_1^2r_2},\qquad
\cos\beta =
   \frac{3r_0^2r_1^2-4r_1^2r_2^2-3r_2^2r_3^2}{4\sqrt{3}r_1r_2^2r_3},
\nonumber \\
 \cos(\alpha+\beta)  =
  - \frac{3r_0^2r_1^2-4r_1^2r_2^2+3r_2^2r_3^2}{6 r_0r_1r_2^2r_3},  \label{eq:cos-choice1}\\
\label{eq:cos-choice2}
 \text{or}\quad \cos\alpha=-1,\qquad \cos\beta =-1,\qquad \cos(\alpha+\beta)  =1,  \\
 \text{or}\quad \cos\alpha=1,\qquad  \cos\beta =-1,\qquad \cos(\alpha+\beta)  =-1,  \\
 \text{or}\quad \cos\alpha=-1,\qquad \cos\beta =1,\qquad \cos(\alpha+\beta)  =-1 .
\label{eq:cos-choice4}
\end{gather}
Inserting \eqref{eq:cos-choice1}, we get
\begin{gather*}
A=\frac{9}{4} -  \left( 3r_0^2+2r_1^2+r_2^2 -\frac{3}{2} \right)^2 ,
\end{gather*}
and thus we f\/ind
\begin{gather}
\label{eq:3r+2r+r}
3r_0^{\blt 2}+2r_1^{\blt 2}+r_2^{\blt 2} =\frac{3}{2} ,
\qquad
  |d|^2 = \frac{1}{4} \lambda_0^\ell
   +\frac{1}{4} \lambda_2^\ell + \frac{1}{2}  \lambda_3^\ell .
\end{gather}
We end up achieving the same value $|d|^2$ \eqref{eq:3r+2r+r}
for \eqref{eq:cos-choice2}--\eqref{eq:cos-choice4}.
For example, inserting~\eqref{eq:cos-choice2}, we get
\begin{gather*}
A=\frac{9}{4} -  \left( 3r_0^2+2r_1^2+r_2^2 -\frac{3}{2} \right)^2
  -  \big(\sqrt{3} r_0r_1-   r_1 r_2 +   \sqrt{3} r_2r_3 \big)^2.
\end{gather*}

\begin{figure}[t]\centering

 \includegraphics[width=75mm]{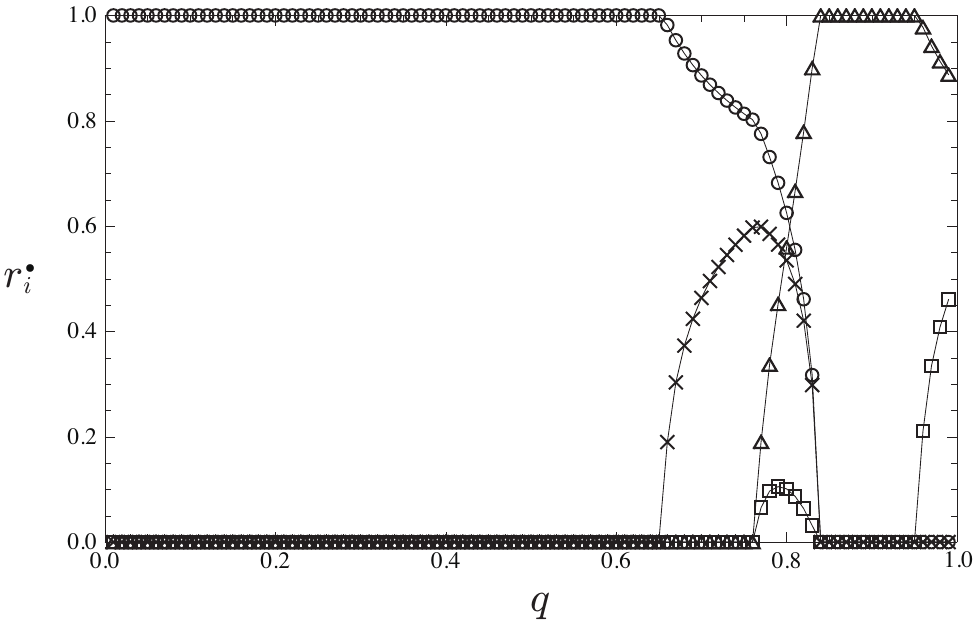}
\qquad
 \includegraphics[width=75mm]{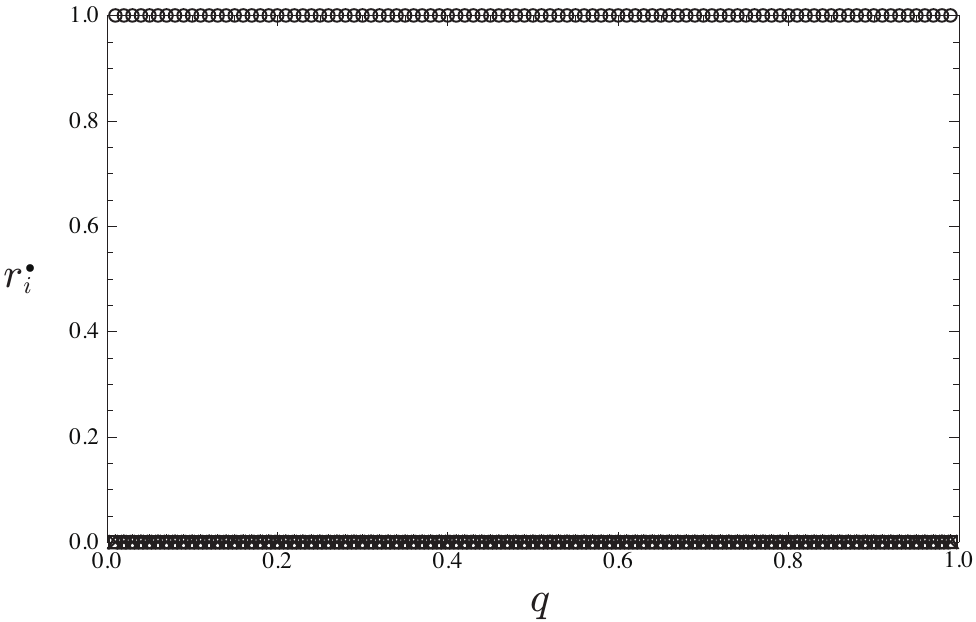}

\caption{Numerical calculations of
the sets $\{r_i^\bullet\}$
which maximize $|d|^2$ for $S=3$ and $\ell=1$ (left)
and~2 (right).
The markers $\bigcirc$, $\triangle$,  $\times$ and $\square$
correspond to $r_0^\bullet$, $r_1^\bullet$, $r_2^\bullet$
 and $r_3^\bullet$, respectively.}\label{fig:S3-ell12}
\end{figure}

The maximization for the anisotropic case
with $\ell$ odd
is more complicated than $S=2$,
see the numerical result   in Fig.~\ref{fig:S3-ell12}.
We expect that
\begin{alignat*}{3}
& \ell \ {\rm odd}: \quad && \text{there exist} \  q^\blt, \ q^{\blt\blt}, \ q^{\blt\blt\blt} \
  {\rm and} \ q^{\blt\blt\blt\blt} \ \text{such that} &  \\
&&& \begin{cases}r^\blt_i=\delta_{i0}, & 0<q\le q^\blt  ,\\
  r_1^\blt=r_3^\blt=0, \ r_0^\blt>0, \ r_2^\blt>0, &   q^\blt<q\le q^{\blt\blt}  , \\
    r_i^\blt>0,\  i=0,1,2,3 , &  q^{\blt\blt}<q<q^{\blt\blt\blt }  , \\
   r_i^\blt=\delta_{i1}, &    q^{\blt\blt\blt}\le q<q^{\blt\blt\blt\blt }  , \\
   r_0^\blt=r_2^\blt=0,\ r_1^\blt>0, \ r_3^\blt>0, &  q^{\blt\blt\blt\blt}\le q<1  ,
    \end{cases} &\\
& \ell \ {\rm even}: \quad &&    r_i^\blt = \delta_{i0} \
     \text{always maximizes} \ \langle \Aux| G^\ell| \Aux \rangle . &
\end{alignat*}
We also expect that these transition points
 $q^\blt,\dots,q^{\blt\blt\blt\blt}$
approach~1 as~$\ell\to\infty$.
Under the assumption $r_i^\blt =\delta_{i0}$, we have
\begin{gather*}
   \mathcal{E}(\ell)  = - \Log \frac{ |d|^2 }{\lambda_0^\ell}
   = -  \Log  \langle\Aux| G^\ell |\Aux\rangle |_{r_i=\delta_{i0}}  \\
 \hphantom{\mathcal{E}(\ell)}{} = \Log q^3[4]
  - \Log  \left( 1+ \frac{q^2 [3]^2 \kappa_1^\ell}{[5]}
      + \frac{q^8   \kappa_2^\ell}{1-q^2+q^4}
      + \frac{q^{14}\kappa_3^\ell}{(1-q^2+q^4)[5]}  \right)  ,
\end{gather*}
for $0<q<1$ and suf\/f\/iciently large $\ell$.

\subsection{General case}

\begin{figure}[t]\centering

\includegraphics[width=90mm]{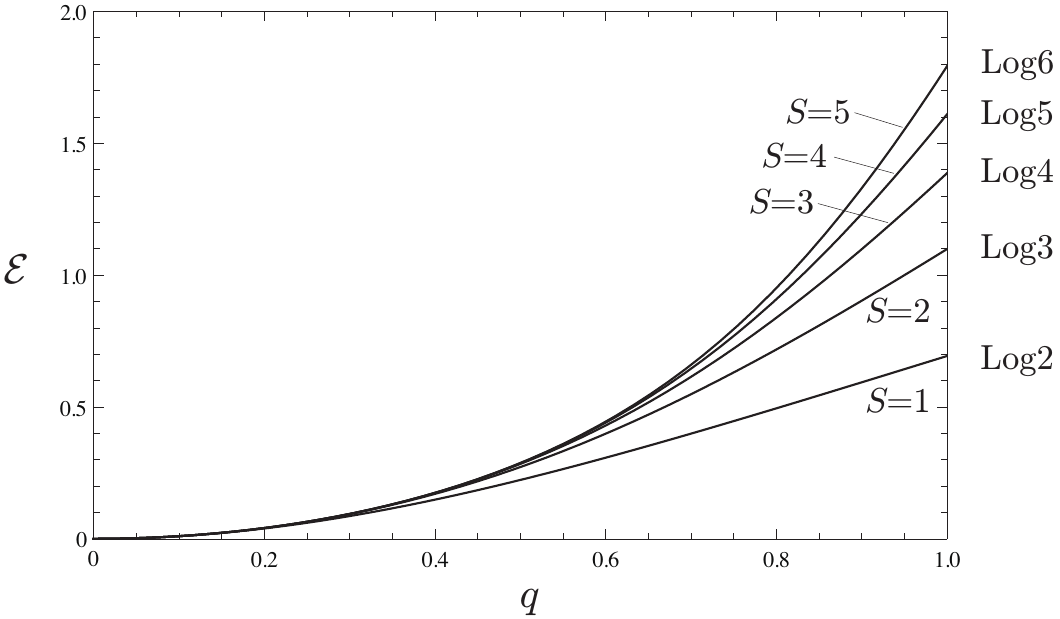}

\caption{The geometric entanglement in the limit $\ell\to\infty$.}
\label{fig:GE-DL}

\end{figure}

We consider the maximization of $\langle \Aux| G^\ell |  \Aux  \rangle $
for general $S$.
As we observed in the previous subsections,
we expect that, for given $q<1$,
\begin{itemize}\itemsep=0pt
  \item[]$\ell$ odd: there exists $\ell^\blt$  such that
    the set $\{r^\blt_i=\delta_{i0}\}$
     maximizes $\langle \Aux| G^\ell| \Aux \rangle$
      when $\ell>\ell^\blt$,
  \item[]$\ell$ even: the set $\{r^\blt_i=\delta_{i0}\}$
     always maximizes $\langle \Aux| G^\ell| \Aux \rangle$.
\end{itemize}
Since the term
$\frac{\lambda_0^\ell}{ _{0}\brabra \lambda_0|\lambda_0 \ketket_{0} }
\left| \sum\limits_{a=0}^{S} q^a r_a^2  \right|^2 $
 dominates in  $\langle \Aux| G^\ell| \Aux \rangle$
for $\ell\to\infty$, we have $r_i^\blt\to \delta_{i0}$,
which supports the above assumption.
Inserting $\{r_i=\delta_{i0}\}$, we have
\begin{gather*}
|d|^2= \bra \operatorname{Aux} |G^\ell| \operatorname{Aux}
\ket \Big|_{r_i =\delta_{i0}}
= \sum_{k=0}^S
\frac{\lambda_k^\ell}{ _{0}\brabra \lambda_k|\lambda_k \ketket_{0} },
\end{gather*}
and f\/ind
\begin{gather}
\mathcal{E}(\ell)=
-\Log \left( \sum_{k=0}^S
\frac{\kappa_k^\ell}{ _{0} \brabra \lambda_k|\lambda_k \ketket_0}
  \right) \nonumber \\
\hphantom{\mathcal{E}(\ell)}{}
   =  \Log (q^S[S+1]) -\Log \left( \sum_{k=0}^S
q^{k(k+1)}
\frac{[2k+1][S]![S+1]!}{[S+k+1]![S-k]!} \kappa_k^\ell
 \right) \nonumber\\
 \hphantom{\mathcal{E}(\ell)}{}
=  \Log (q^S[S+1]) - \frac{q^2[3][S]}{[S+2]} \kappa_1^\ell + o\big( \kappa_1^\ell\big),
   \qquad  \ell\to\infty    .
\label{eq:GE-fini-size-corr}
\end{gather}
Here we used the norm~\eqref{norm}
of the $q$-deformed VBS state $|\Psi \rangle$
and the norm of the eigenvectors of the transfer matrix  \cite{AM}.
In the limit $\ell\to\infty$, we have
the geometric entanglement  $\mathcal{E} = \Log (q^S[S+1]) $,
which takes the maximum $\Log(S+1)$ at $q=1$
 and approaches 0 as $q\to0$,
see Fig.~\ref{fig:GE-DL}.
The monotonic behavior while $0<q<1$ is similar to the entanglement entropies.

The isotropic point is a special case
where the choice $r_i=\delta_{0i}$ or $r_i=\delta_{Si}$
does not always maximize $|d|^2$ for $S\ge 2$
even if $\ell$ is large,
as we saw for $S=2$ and $S=3$.
Thus the  asymptotic form~\eqref{eq:GE-fini-size-corr}
is no longer  valid at the isotropic point.

\section{Summary and discussion}\label{section5}

In this article, we studied some entanglement properties
of the higher spin $q$-AKLT model with the periodic boundary condition
from the matrix product representation of the $q$-VBS ground state.
We exactly calculated the f\/inite-size correction terms
of the entanglement entropies by the perturbative calculation
for the spectrum of the reduced density matrix.
We found that  the f\/irst-order correction term
of the R\'enyi entropy vanishes
by taking the limit $\alpha\to 1$.
This requires the second-order perturbation of the entanglement spectrum
for calculation of the f\/inite-size correction of the von Neumann entropy.
It would be interesting to extend the study of entanglement
properties to various generalizations,
the entanglement entropies with multiple blocks
(see~\cite{SK} for the isotropic spin-1 case), for example.
We also investigated the geometric entanglement.
The geometric entanglement in the limit $\ell\to\infty$
decreases with the decrease of the anisotropy parameter $q$
while $0<q<1$.
This property is the same as the entanglement entropies.
Under an assumption which is based on numerical results,
we calculated the f\/inite-size correction of the geometric entanglement.

\subsection*{Acknowledgements}

C.~Arita is a JSPS Fellow for Research Abroad.

\pdfbookmark[1]{References}{ref}
\LastPageEnding

\end{document}